\documentclass[twocolumn]{aastex62}
\hypersetup{breaklinks}
\usepackage{comment}

\newcommand{\caiihk}{\ion{Ca}{2} H \& K}

\newcommand{\teff}{\ensuremath{T_{\mbox{\scriptsize eff}}}}

\newcommand{\msun}{\ensuremath{\mbox{M}_{\odot}}}

\newcommand{\mjup}{\ensuremath{\mbox{M}_{\rm Jup}}}

\newcommand{\prot}{\ensuremath{P_{\mbox{\scriptsize rot}}}}

\newcommand{\kms}{\ensuremath{\mbox{km s}^{-1}}}

\newcommand{\degree}{\ensuremath{^\circ}}
\newcommand{\mas}{\ensuremath{\mbox{mas yr}^{-1}}}

\newcommand{\gbr}{\ensuremath{(G_{\rm BP} - G_{\rm RP})}}

\newcommand{\stream}{Psc--Eri stream}

\accepted{May 24, 2019}
\submitjournal{\textit{The Astronomical Journal}}

\shorttitle{\sc \textit{TESS} Reveals that the Nearby Pisces--Eridanus Stellar Stream is only 120 Myr Old}
\shortauthors{Curtis et al.}

\begin{document}

\title{\sc \textit{TESS} Reveals that the Nearby Pisces--Eridanus Stellar Stream is only 120 Myr Old}

\newcommand{\cehw}{Center for Exoplanets and Habitable Worlds, Department of Astronomy \& Astrophysics, 
    The Pennsylvania State University, \\ 
    525 Davey Laboratory, University Park, PA 16802, USA}
\newcommand{\columbia}{Department of Astronomy, Columbia University, 550 West 120th Street, New York, NY 10027, USA}
\newcommand{\jpl}{Jet Propulsion Laboratory, California Institute of Technology, M/S
  321-100, 4800 Oak Grove Drive, Pasadena, CA 91109, USA}
\newcommand{\rochester}{Department of Physics \& Astronomy, University of
  Rochester, Rochester, NY 14627, USA}

\newcommand{\jhu}{Center for Astrophysical Sciences, Johns Hopkins University,
3400 N. Charles Street, Baltimore, MD 21218, USA}

\correspondingauthor{Jason Lee Curtis}
\email{jasoncurtis.astro@gmail.com}

\author[0000-0002-2792-134X]{Jason L.~Curtis}
\altaffiliation{NSF Astronomy and Astrophysics Postdoctoral Fellow}
\affiliation{\columbia}

\author[0000-0001-7077-3664]{Marcel A.~Ag\"{u}eros}
\affiliation{\columbia}

\author[0000-0003-2008-1488]{Eric E. Mamajek}
\affiliation{\jpl}
\affiliation{\rochester}

\author[0000-0001-6160-5888]{Jason T. Wright}
\affiliation{\cehw}

\author[0000-0001-7453-9947]{Jeffrey D. Cummings}
\affiliation{\jhu}



\begin{abstract}
Pisces--Eridanus (Psc--Eri), a nearby ($d$ $\simeq$ 80-226 pc) stellar stream stretching across $\approx$120\degree\ of the sky, was recently discovered with  \textit{Gaia} data.
The stream was claimed to be $\approx$1~Gyr old, which would make it an exceptional  discovery for stellar astrophysics, as star clusters of that age are rare and tend to be distant, limiting their utility as benchmark samples. 
We test this old age for Psc--Eri in two ways. First,  we compare the rotation periods for 101 low-mass members (measured using time series photometry from the \textit{Transiting Exoplanet Survey Satellite}, \textit{TESS}) to those of well-studied open clusters. Second, we identify 34 new high-mass candidate members, 
including the notable stars $\lambda$~Tauri (an Algol-type eclipsing binary) and HD~1160 (host to a directly imaged object near the hydrogen-burning limit).
We conduct an isochronal analysis of the color--magnitude data for these highest-mass members, again comparing our results to those for open clusters.
Both analyses show that the stream has an age consistent with that of the
Pleiades, i.e., $\approx$120 Myr. 
This makes the Psc--Eri stream an exciting source of 
young benchmarkable stars and, potentially, exoplanets located in a more diffuse environment that is distinct from that of the 
Pleiades and of other dense star clusters.
\end{abstract}




\section{Introduction} \label{s:intro}

\begin{figure}\begin{center}
\includegraphics[trim=0.95cm 0cm 0.0cm 0.7cm, clip=True,  width=3.4in]{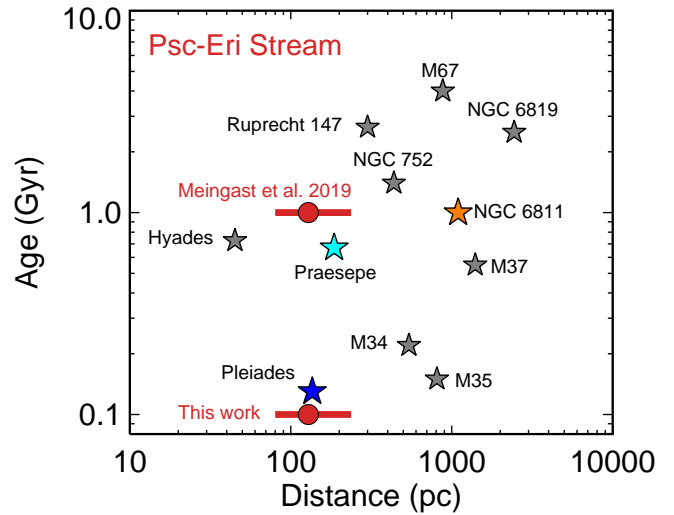}
  \caption{Age versus distance for a selection of benchmark star clusters with rotation period measurements.
  The distance to the \stream\ is shown as a red point marking the median and a red line   showing the range. If this stream is really $\approx$1~Gyr in age, it would become a critical target for rotation/activity studies and an important benchmark for stellar astrophysics. By comparing rotation periods in Psc--Eri to those in the clusters shown as colored stars, and by re-examining its color-magnitude diagram, we demonstrate that it is closer to $\sim$100~Myr old.
    \label{f:cfam}}
\vspace{-0.6cm}
\end{center}\end{figure}

Star clusters at least 1~Gyr in age are rare, and tend to be located at large distances from Earth \citep[e.g.,][]{Dias2002,khar2005}.
This is a shame, because such clusters serve as critical benchmarks for stellar astrophysics. Recently, \citet{Meingast2019} announced the discovery of a stellar stream 
that stretches 120\degree\ across the sky, and spans $\approx$400~pc in space. 
This discovery was made possible by the precise astrometry, 
radial velocities (RVs), and photometry included in the \textit{Gaia} mission's second data release \citep[DR2;][]{GaiaDR2}.
Discovery of the Pisces--Eridanus stream (Psc--Eri)\footnote{The stream was undesignated
in \citet{Meingast2019}.  
The authors of the discovery paper suggested the name ``MAF-1'' for the stream (S.~Meingast, priv.~comm.); however, this is very different from the nomenclature for nearby associations
\citep[e.g.][]{deZeeuw1999, Torres2008}. This acronym could be confused with two acronyms already in the Dictionary of Nomenclature of Celestial Objects \citep[\url{http://cds.u-strasbg.fr/cgi-bin/Dic-Simbad};][]{Lortet1994}---[MAF2004] and [MAF2009]---the latter of which is used for members of the open cluster NGC 7062
\citep{Molenda-Zakowicz2009}, or as an abbreviation of the Maffei  galaxies
or Maffei Group of galaxies \citep[e.g.][]{Fingerhut2007}. Two of the main concentrations of the stream's members are in the constellations Eridanus (clump 1) and Pisces (clump 3), and the group's convergent point 
($\alpha, \delta$ $\simeq$ 42\fdg6, $-$20\fdg0; ICRS) lies in Eridanus as well. As we find in our analysis that the group is more analogous to an older version of an OB association, similar to other expansive nearby stellar associations like Sco--Cen and Tuc--Hor, we combine the two prominent constellation names and refer to it as the ``Pisces--Eridanus stream" or Psc--Eri.
}
was somewhat of a surprise
given its combination of old age ($\approx$1~Gyr) and proximity ($d = 129$$\pm$$32$~pc from Earth; median and standard deviation of the 256 published members; the full range is $d$ $\simeq$ 80--226~pc). 
For context, we list the distance moduli for notable benchmark open clusters along with their ages in Table~\ref{t:clus}. Figure~\ref{f:cfam} plots the age and distance to a selection of  clusters with measured rotation periods (\prot), which further highlights how remarkable and useful a 1~Gyr cluster 
this close to Earth would be.

\begin{deluxetable}{lccl}
\tablecaption{Ages and distance moduli for notable benchmark star clusters with rotation data.} \label{t:clus}
\tablehead{\colhead{Name} & \colhead{$m - M$} & \colhead{Age (Gyr)} & \colhead{Age Reference}
}
\startdata
Pleiades &  5.67 & 0.120 & \citet{Stauffer1998}\tablenotemark{b} \\
Praesepe &  6.35 & 0.670 & \citet{Douglas2019} \\
Hyades   &  3.37 & 0.730 & \citet{Douglas2019} \\
NGC~6811 & 10.20 & 1.0 & \citet{Curtis2019} \\
NGC~752  &  8.20 & 1.4 & \citet{Agueros2018} \\
Ruprecht~147 & 7.40 & 2.7 & \citet{Torres2018} \\
M67 & 9.72 & 4.0 & \citet{M67SolarTwin} \\
\enddata
\tablecomments{These distance moduli only account for 
distance, and do not include visual extinction.}
\tablenotetext{a}{The Pleiades age has been constrained with lithium depletion boundary 
to 125--130~Myr by \citet{Stauffer1998} and $115 \pm 5$~Myr by \citet{Dahm2015}.
Recent isochrone analyses by \citet{Gossage2018} found 110--160~Myr; 
\citet{Cummings2018young} found 115--135~Myr. We adopt 120~Myr for this work.}
\end{deluxetable}

If Psc--Eri's age is truly 1 Gyr, it would be the oldest 
coeval stellar population within 300~pc.
This would open up many avenues for research that are difficult or impossible to pursue with the 1 Gyr-old benchmark cluster NGC 6811 \citep{Sandquist2016,Curtis2019}, 
currently the only open cluster of this age we have been able to study in detail. 
For example, \citet{meibom2013} discovered two 
sub-Neptune exoplanets in NGC 6811, but 
these are too faint for efficient RV follow-up. It is also challenging to measure chromospheric \caiihk\ activity indices for FGK stars in NGC 6811: those stars are faint (a solar twin is $V \approx 15$), and the interstellar \caiihk\ contamination is difficult to mitigate \citep{Curtis2017}. 
Finally, Psc--Eri could be an interesting test case for demonstrating the chemical tagging technique needed for Galactic archaeology  \citep{fbh2002}.

Given the potential value of this population of stars, it is important to examine its age to see if it can serve as a benchmark for old stars. A similar exercise with the purportedly old nearby cluster Ruprecht 147 proved very fruitful \citep{Curtis2013, Curtis2016PhD}, while the exploration of another candidate old cluster, Lod\'en 1, showed that it did not exist \citep{LodenPaper}.

We use gyrochronology, the age-dating method based on stellar rotation and magnetic braking \citep{Barnes2003, soderblom2010}, to test the existence and coevality of the \stream. Coeval stars form well-defined sequences in their color--period diagrams, analogous to the main sequence in a color--magnitude diagram (CMD). But color--period sequences are much more sensitive to age, as the full sequence evolves measurably in time as stars spin down, while only the massive end of the main sequence shows significant evolution in temperature and luminosity.
If the stars are coeval, a gyrochronology analysis will also yield a precise age for the \stream. We conduct this experiment in  Section~\ref{s:gyro}, where we extract and analyze light curves for 101 members of the stream observed by \textit{TESS}. We find that the resulting \prot\ distribution precisely overlaps the Pleiades distribution, making it  $\approx$120~Myr old.

In Section~\ref{s:cmd}, we reinterpret the stream's CMD by noting that \textit{Gaia} DR2 measured RVs for stars with $\teff \lesssim 7000$~K, which biased the \citet{Meingast2019} membership census. 
The \stream's CMD closely 
matches that of the Pleiades,  
except that its membership is truncated due to this RV bias. Combining \textit{Gaia} DR2 data
with literature RVs, we identify 22 new candidates that are warmer than the stars in the \citet{Meingast2019} sample, and another 12 that lack RVs but are co-moving in proper motion within 10~pc of known members. These stars closely follow the upper main sequence of the Pleiades, providing further evidence of the \stream's young age. We also briefly discuss the stream's formation in Section~\ref{s:cmd}, before  concluding in Section~\ref{s:end}.

\section{Age-dating the Psc--Eri Stream with Gyrochronology} \label{s:gyro}

\subsection{Rotation Period Measurements with \textit{TESS}} \label{s:lc}

The \textit{Transiting Exoplanet Survey Satellite} \citep[\textit{TESS};][]{TESS} is currently conducting a year-long photometric monitoring campaign of the southern sky. \textit{TESS} scans the sky in a series of sectors for $\approx$27~d at a time. Full frame images (FFI) are recorded with a 30~m cadence. As of writing, FFI data for the first five sectors have been released to the Mikulski Archive for Space Telescopes (MAST). 

\citet{Meingast2019} published a list of 256 candidates members of the \stream. We used the Web \textit{TESS} Viewing Tool (WTV)\footnote{\url{https://heasarc.gsfc.nasa.gov/cgi-bin/tess/webtess/wtv.py}} 
to identify stars observed during Sectors 1-5, 
and we found 154 with data from at least one sector.
We downloaded 20$\times$20 pixel cutouts of the FFI images 
centered on each target using the 
\texttt{TESSut} tool hosted at MAST \citep{Astrocut}.\footnote{\url{https://mast.stsci.edu/tesscut/}}
Next, we used the IDL procedure \texttt{aper.pro} 
from the IDL Astronomy User's Library 
\citep{IDLastro} to perform aperture photometry
on all epochs in the image stack produced by 
\texttt{TESScut}.
We used a circular aperture with a 3 pixel radius
($\approx$1$'$ based on \textit{TESS}'s $\approx$21$''$ pixel scale).

\begin{figure*}\begin{center}
\includegraphics[trim=0.0cm 0cm 0.0cm 0cm, clip=True,  width=6.5in]{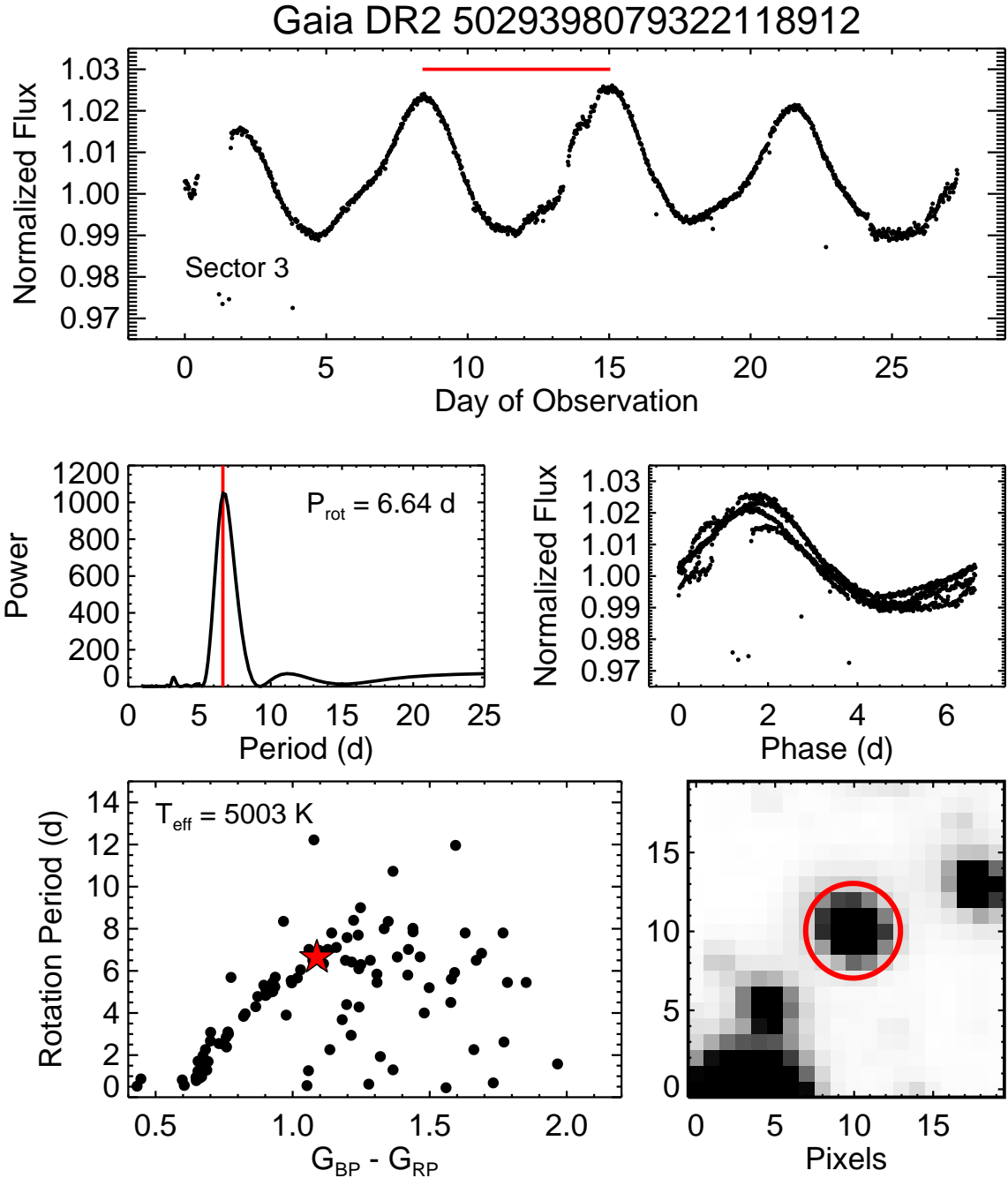}
  \caption{\textit{Top---}Example \textit{TESS} light curve for 
   Gaia DR2 5029398079322118912, which was observed during Sector 3.
   The length of the red line at the top left is the duration of one cycle (i.e., \prot).  
  \textit{Middle left---}The Lomb--Scargle periodogram shows 
  $\prot = 6.64$~d. 
  In some cases, the periodogram did not produce an accurate measurement, 
  so we calculated \prot\ by fitting the timing of successive maxima and/or minima, 
  illustrated by the red line in the top panel. 
  \textit{Middle right---}This phase-folded light curve visually validates the periodogram analysis.
  \textit{Bottom left---}The color and period for this star (red star)
  are plotted along with the full rotator sample for the \stream\ (black points).
    The \textit{Gaia} DR2 \teff\ is also provided \citep{DR2prop}.
  \textit{Bottom right---}The 20$\times$20 pixel cutout of the \textit{TESS} full frame image for this target, encircled with a 
  three~pixel radius aperture used to extract the light curve (red circle). 
  Versions of this figure for every target analyzed are available as an electronic figure set in the online Journal (101 images) . 
    \label{f:lcs}}
\end{center}\end{figure*}

\begin{figure*}\begin{center}
\includegraphics[trim=0.0cm 0cm 0.0cm 0cm, clip=True,  width=6.5in]{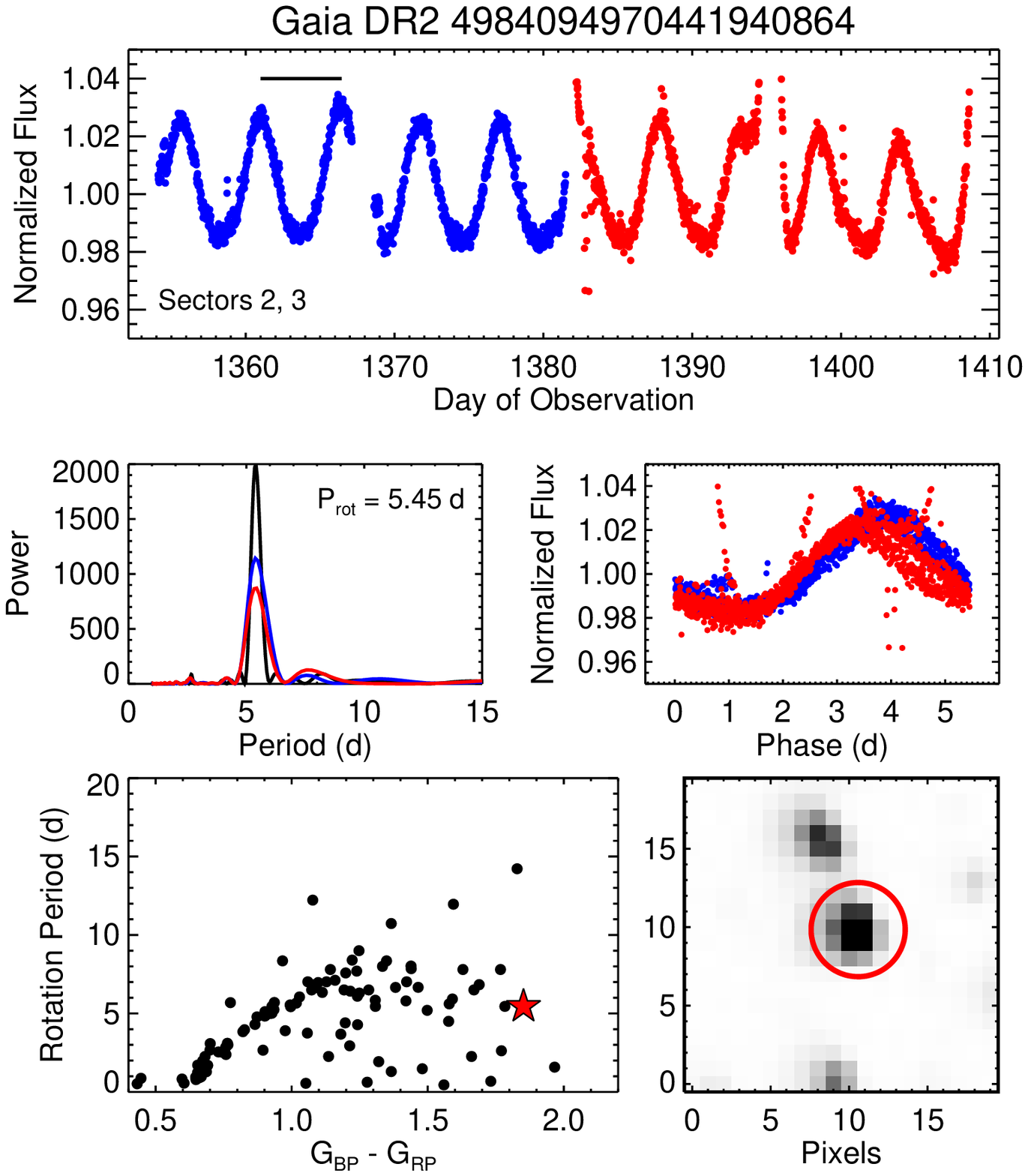}
  \caption{Similar to Figure~\ref{f:lcs}, 
  but for the 11 stars with two sectors of data. \textit{Top---}The \textit{TESS} light curve for 
  Gaia DR2 4984094970441940864, which was observed during Sectors 2 (blue) and 3 (red).
  The length of the black line at the top left is the duration of one cycle (i.e., \prot).  
  \textit{Middle left---}Lomb--Scargle periodograms 
  for Sector 2 (blue), Sector 3 (red), and the joint light curve (black). While we find the same period from each individual sector, the periodogram peak
  is noticeably narrower for the joint light curve.
  \textit{Middle right---}The phase-folded light curves for each sector show that the light curves can be reliably merged by simply stitching them together with no additional calibration needed (for these rapid, active stars, at least). 
  \textit{Bottom left---}The color and period for this star (red star)
  are plotted along with the full rotator sample for the \stream\ (black points).
  \textit{Bottom right---}The 20$\times$20 pixel cutout of the \textit{TESS} full frame image for this target, encircled with a three~pixel aperture used to extract the light curve (red circle).
     \label{f:multi}}
\end{center}\end{figure*}

The resulting light curves overwhelmingly showed clear spot modulation with relatively large amplitudes and short periods compared to our expectations from the 1 Gyr NGC 6811 data from \textit{Kepler} \citep{Curtis2019, Meibom2011}. We were able to measure \prot\ without performing any additional calibration on these light curves. Figures \ref{f:lcs} and \ref{f:multi} show examples of \textit{TESS} light curves for stream members
produced following this simple procedure.


\subsection{The Color--Period Diagram} \label{s:cpd}
We measured rotation periods for 101 stars 
using Lomb--Scargle periodograms
\citep{Scargle1982, press1989}.
After extracting each light curve and computing the periodogram, we visually inspected the results (see Figures \ref{f:lcs} and \ref{f:multi}) to ensure the accuracy of our measurements.
On only three occasions did we double the Lomb--Scargle period to correct for a 1/2-period harmonic error, which we visually identified by noticing asymmetry in the depths of alternating minima and other subtle morphological asymmetries.

Eleven stars were observed twice, in neighboring sectors, and for these we find consistent periods across sectors. Figure~\ref{f:multi} shows an example where we stitched the light curves from two sectors together, 
and found a more precise period than attained from either sector separately (based on the width of the periodogram peak).
Stitching the light curves together was simplified by the fact that multiple maxima and minima were captured in each sector, 
which meant that no reference stars were needed to normalize the light curves from each sector.

The bottom left panels of Figures~\ref{f:lcs} and \ref{f:multi} plot \textit{Gaia} DR2 color \gbr\ versus \prot\ for our sample. The majority of the stars follow a common sequence, indicating that they are coeval. 



\begin{figure*}\begin{center}
\includegraphics[trim=1.5cm 0cm 0.3cm 0cm, clip=True,  width=3.4in]{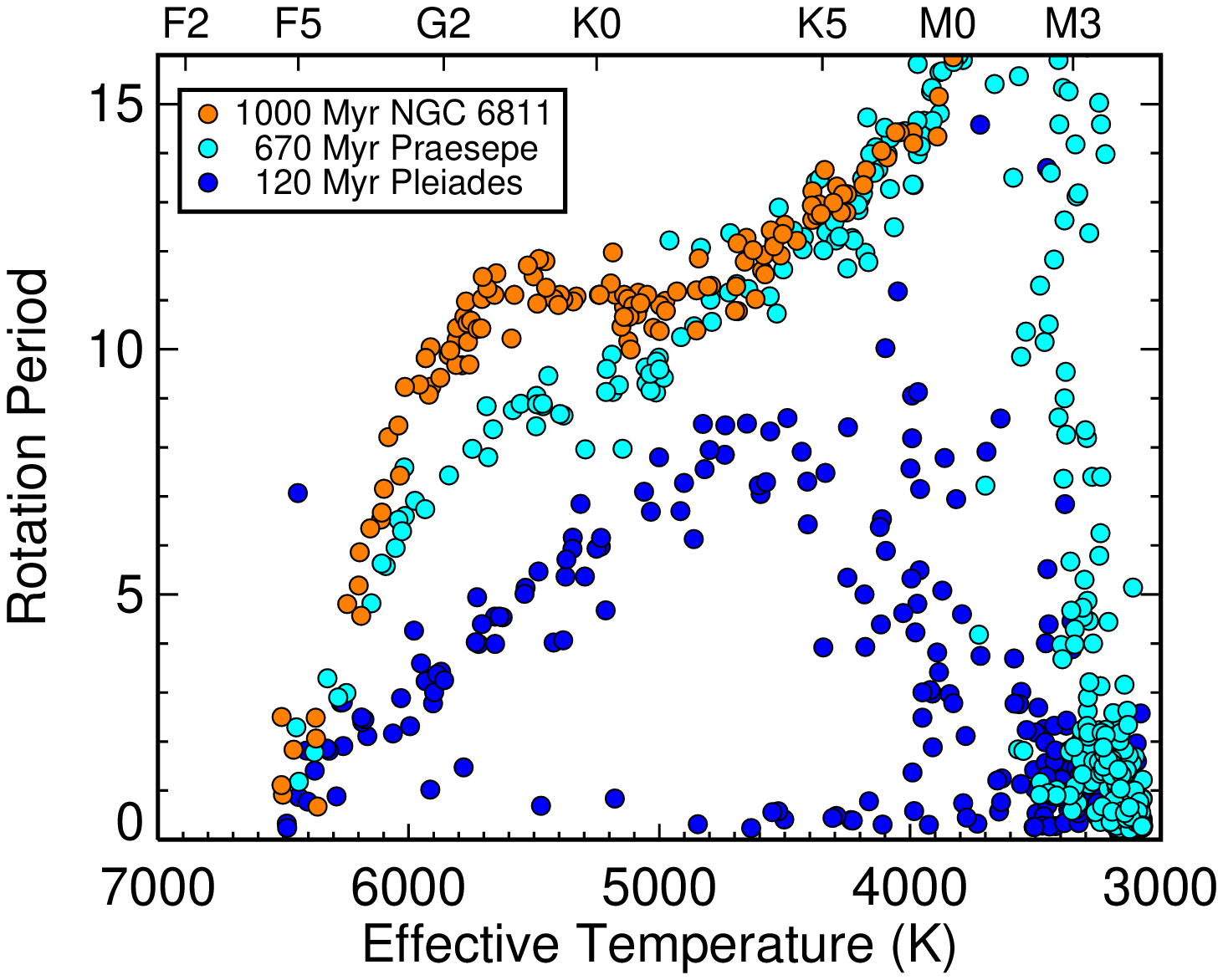}
\includegraphics[trim=1.5cm 0cm 0.3cm 0cm, clip=True,  width=3.4in]{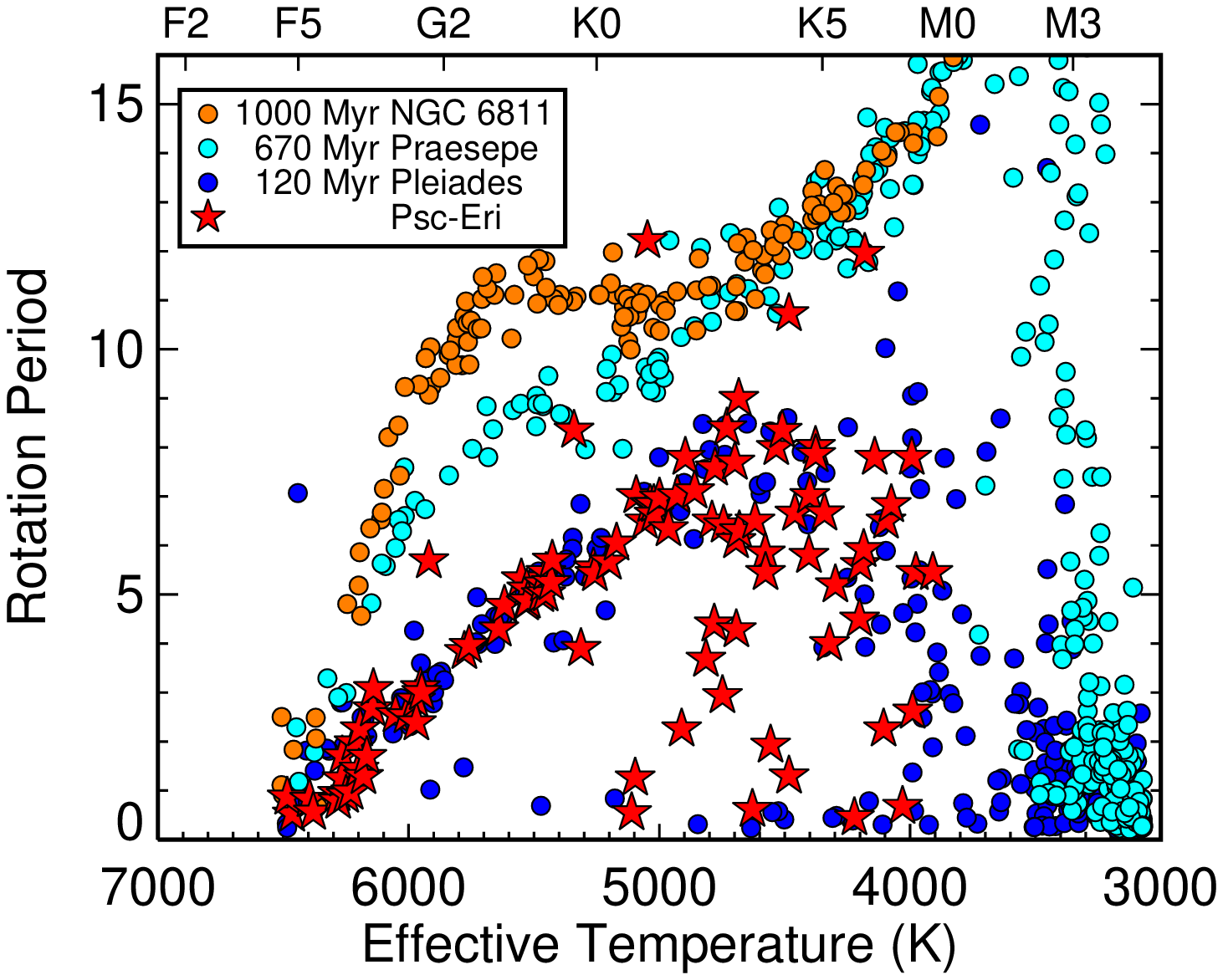}
  \caption{\textit{Left}---Rotation period distributions for single-star members of 
  the Pleiades \citep[blue points, 120 Myr;][]{Rebull2016}, 
  Praesepe \citep[cyan points, 670 Myr;][]{Douglas2017,Douglas2019}, and 
  NGC~6811 \citep[orange points, 1 Gyr;][]{Curtis2019}.
  \textit{Right}---Similar to the previous panel, and now including rotation periods for 
  101 members of the \stream\ (red stars)
  identified by \citet{Meingast2019}, 
  and measured by us from \textit{TESS} FFI data. Approximate spectral types are listed at the top of each panel for reference.
  Clearly, the rotation period distribution 
  for this stream favors an age much younger than 
  1 Gyr. We infer an age of $\approx$120~Myr 
  for the \stream\ based on its similarity with the Pleiades.  
    \label{f:gyro}}
\end{center}\end{figure*}

\subsection{A Gyrochronological Age} \label{s:gage}
Gyrochronology only requires as input the mass of a star (or a proxy like temperature or color) and its \prot. There are a variety of empirical gyrochronology models available, including those of \citet{Barnes2003}, \citet{barnes2007} and its various re-calibrations \citep[e.g.,][]{mamajek2008, Angus2015}, and \citet{Barnes2010}. 
There are also theoretical models that pair stellar evolution with a magnetic torque law to predict angular momentum evolution \citep[e.g.,][]{vanSaders2013, Matt2015, Gallet2015}. However, no model has been published that can explain all of the cluster rotation data \citep[see][]{Curtis2019, Douglas2019, Agueros2018}.
Instead, we suggest that the best way to constrain the age of the \stream\ with gyrochronology is by comparing its \prot\ distribution directly to the distributions measured for benchmark clusters.

\subsubsection{The Benchmark Cluster Sample}

The Pleiades is $\approx$120 Myr old \citep[][see also Table~\ref{t:clus}]{Stauffer1998}, 
has a metallicity of [Fe/H] = +0.03 dex \citep{Soderblom2009} and
an interstellar reddening of $E(B - V) \approx 0.044$ \citep[$A_V = 0.14$;][]{Taylor2006}. 
\prot\ for 759 members were measured by \citet{Rebull2016} from \textit{K2} light curves collected during its Campaign~4 \citep[see also][]{Rebull2016b,Stauffer2016}. We cross-matched this list with \textit{Gaia} DR2 and filtered out stars that were more than 0.375 mag discrepant from the single-star sequence, which we defined with the \citet{DR2HRD} membership list; this is half of the offset for an equal-mass binary \citep[e.g.,][]{hodgkin1999}. We also removed stars with absolute differences in proper motion relative to the cluster median greater than 3~\mas, corresponding to $\approx$2~\kms\ at 136~pc, or four times the internal velocity dispersion 
\citep{Madsen2002}.

Praesepe is 670 Myr old \citep{Douglas2019} and has a metallicity of [Fe/H] = +0.15 dex \citep{Cummings2017}. \prot\ for 743 members were amassed from the literature and measured from \textit{K2} Campaign~5 light curves by \citet{Douglas2017}. \citet{Douglas2019} cross-matched this list with DR2 and filtered out stars that failed membership, multiplicity, and data quality criteria, leaving us with 359 single star members.

The 1 Gyr-old NGC 6811 cluster has a solar metallicity \citep{Sandquist2016}. \prot\ for 171 likely single-star members were recently measured by \citet{Curtis2019}, more than doubling the size of the rotator sample from \citet{Meibom2011}, and extending its lower mass limit from $\approx$0.8~\msun\ to $\approx$0.6~\msun.  

\subsubsection{Stellar Properties}

\textit{Gaia} DR2 provided effective temperatures (\teff) for $\approx$1.61$\times$10$^8$ stars with $3000 \lesssim \teff \lesssim 10,000$~K and 
$G < 17$ mag \citep{DR2HRD} via the Apsis pipeline \citep{apsis2013}. The DR2 photometry is very precise, but the Apsis temperatures are severely affected by interstellar reddening. However, this bias can be mitigated by de-reddening the photometry for each cluster sample prior to converting it to \teff. We employ an empirical color--temperature relation to convert the de-reddened \textit{Gaia} DR2 $\gbr_0$ color to \teff. Our relation is a polynomial fit to benchmark stellar data assembled from  the catalog of spectroscopic properties for the solar-type stars ($4700 < \teff < 6700$~K) targeted by the California Planet Survey \citep{Brewer2016}, warmer stars taken from the Hyades \citep{DR2HRD} with \teff\ from the DR2/Apsis pipeline \citep{DR2prop},  and cooler K and M dwarfs from the \citet{Boyajian2012} and \citet{Mann2015} catalogs. We have also applied this relation in \citet{Morris2018}, 
\citet{Douglas2019}, and \citet{Curtis2019}.

\subsubsection{The Psc--Eri Stream is Coeval with the Pleiades}

In the left panel of Figure \ref{f:gyro}, we present the \prot\ distribution for likely single-star members of our three benchmark open clusters as a function of \teff. In the right panel, we add the the \prot\ distribution for the \stream. 
The \stream's \prot\ distribution is nearly indistinguishable from that of the Pleiades. In particular, the slow, converged sequences for each system are remarkably consistent. 

There are a few differences. The \stream\ has more outliers at periods intermediate to the slow sequence and the rapid $\approx$1~d rotators. This could be due to poor binary rejection,
or slight differences in age---if younger than the Pleiades, those stars could still be converging. In addition, the Pleiades sample extends to much cooler \teff. As we discuss in Section~\ref{s:to}, this is because RVs were used to identify members of the \stream, and DR2 does not provide RVs for such cool and faint stars. Finally, the warmest stars in the \stream\ ($\teff \gtrsim 6100$~K) appear to be rotating subtly and systematically faster than their analogs in the Pleiades. Perhap this also indicates that the stream 
is slightly younger than the Pleiades. 

In contrast, the late-F to early-K dwarfs are, again, remarkably consistent.
The slow, converged sequences for both populations are 
well-described by a line of constant Rossby number.\footnote{$Ro = \prot / \tau$. 
We  used the formula for convective turnover time, $\tau$, from \citet{Cranmer2011}.}
Focusing on the stars with $4600 < \teff < 6100$~K that have 
converged to within 25\% of the slow sequence, 
the median and standard deviation of the Rossby number for the 43 Pleiades in this sample is 
$Ro = 0.29 \pm 0.03$, 
and we find $Ro = 0.29 \pm 0.02$ for the 39 stream members meeting the same criteria.
These values are incredibly precise, and strikingly similar.
The unavoidable conclusion is that the \stream\ is $\approx$120~Myr in age.\footnote{We performed similar comparisons with M35 
\citep[NGC~2168, 150~Myr;][]{MeibomM35}
and 
M34 \citep[NGC~1039, 220~Myr;][]{MeibomM34}, 
and found that the Psc--Eri \prot\ distribution was 
most consistent with that of the Pleiades. 
Specifically, the slow sequences for the older M35 and M34 clusters are converged to lower masses and longer periods \citep[see figure~12 in][]{Stauffer2016}, 
whereas the slow sequences for Psc--Eri and the Pleiades share a common maximum \prot\ of $\approx$8.5~d, where the 
distributions turn over toward more rapid rotation toward
lower masses and cooler temperatures.} 

\section{Revisiting the Psc--Eri Stream's CMD} \label{s:cmd}

The left panel of Figure~\ref{f:cmd} is the CMD for the stream,\footnote{We adopt $d = 1000 / \varpi$ to estimate distances for each star, and so calculate absolute magnitudes as $M_G = G - 5\, \log_{10} (100 / \varpi )$, with units of pc and mas for $d$ and $\varpi$.}

together with members of the Pleiades \citep{DR2HRD} 
and NGC~6811 \citet{Curtis2019}.
We also include PARSEC isochrones \citep{parsec} appropriate for the Pleiades (130~Myr, solar metallicity), and NGC~6811 (1~Gyr, solar metallicity).

\subsection{The Apparent Absence of a Main-Sequence Turnoff Is a Problem} \label{s:to}

\begin{figure*}\begin{center}
\includegraphics[trim=1.0cm 0cm 0.0cm 0cm, clip=True,  width=3.4in]{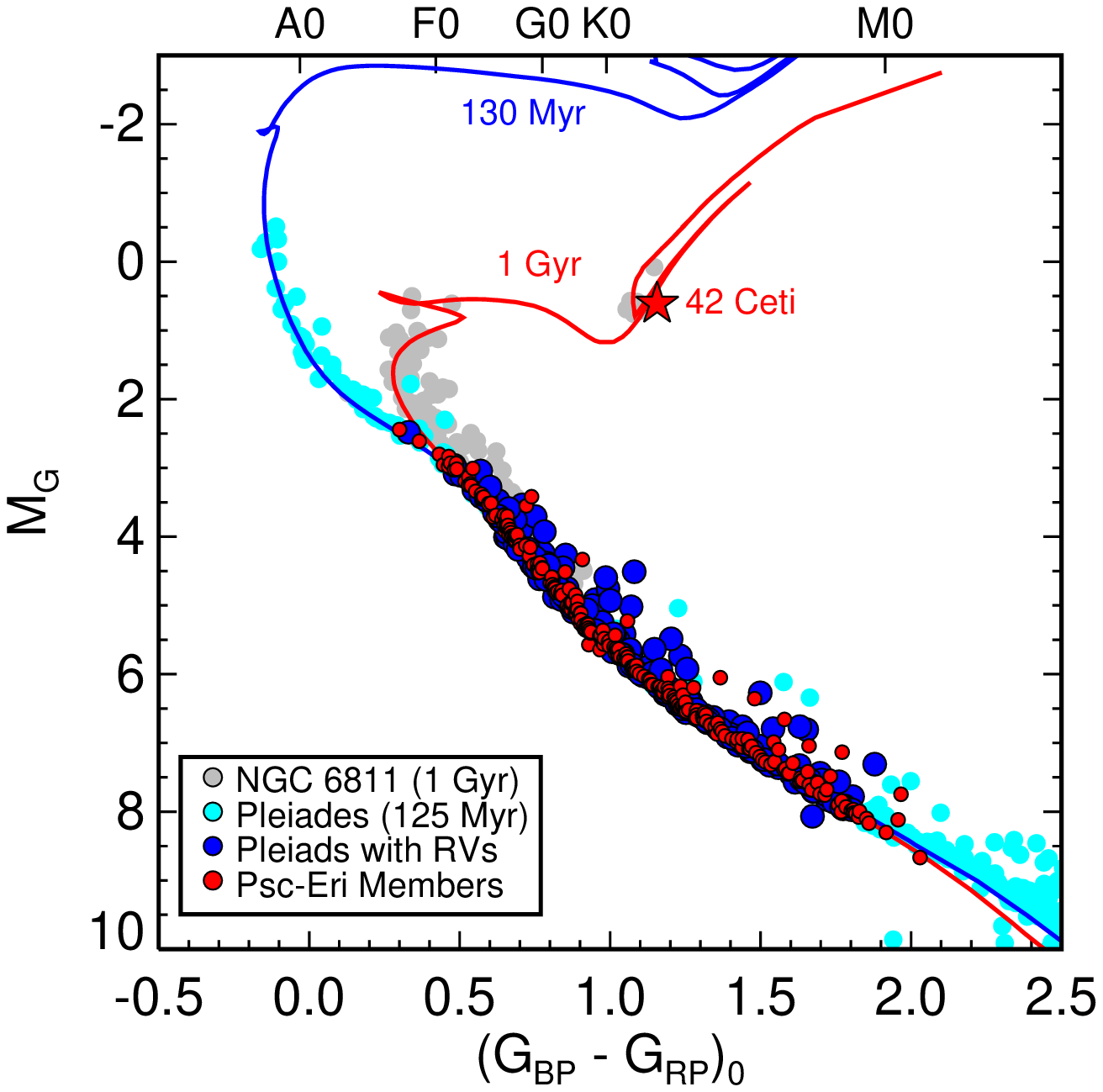}
\includegraphics[trim=1.0cm 0cm 0.0cm 0cm, clip=True,  width=3.4in]{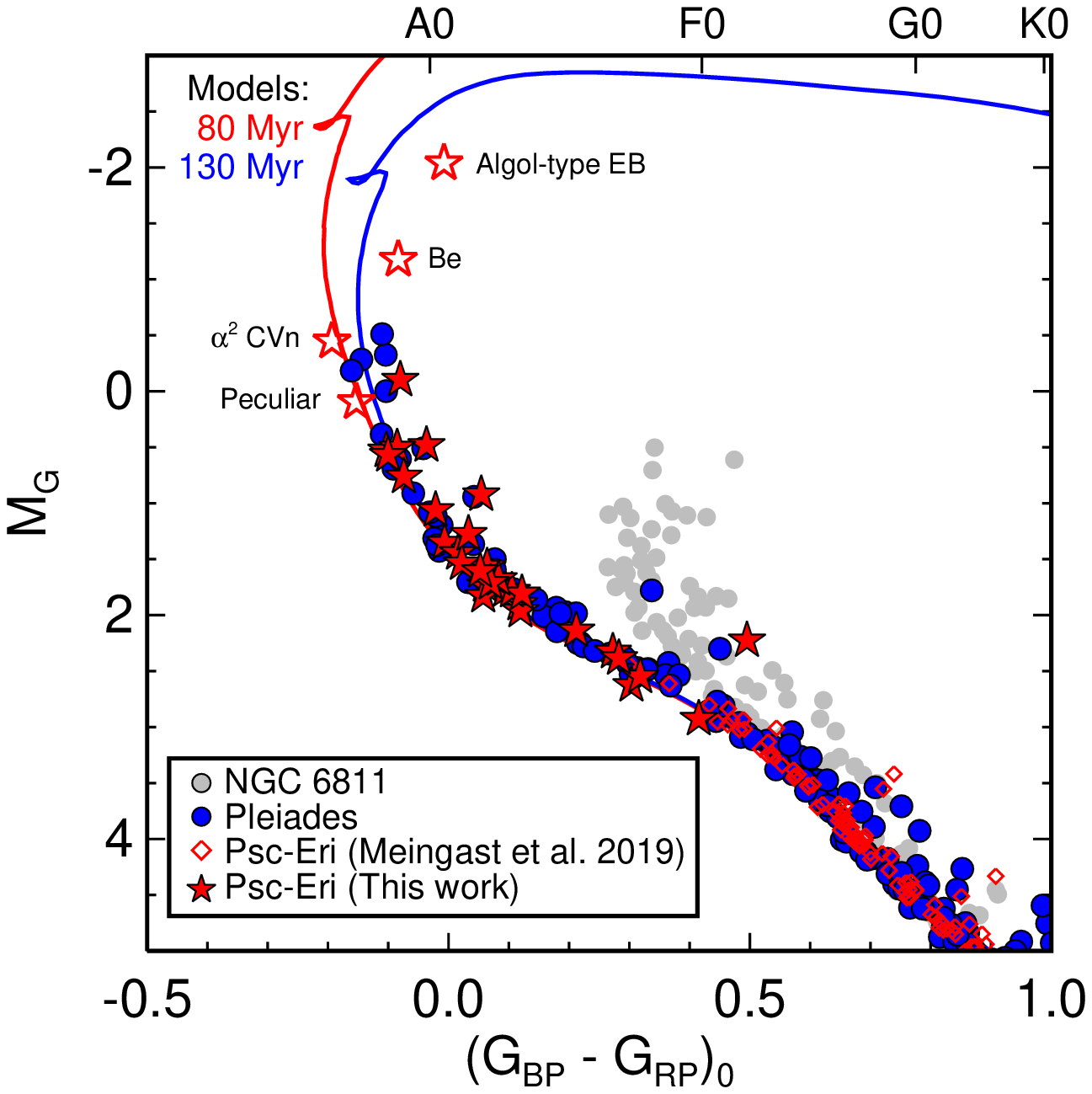}
  \caption{\textit{left---}De-reddened \textit{Gaia} DR2 color versus absolute magnitude for 
  NGC 6811 \citep[gray points; $(m - M)$ = 10.2, $A_V = 0.15$;][]{Curtis2019},
  the stream \citep[red points; distance moduli are calculated directly from parallax, zero reddening is assumed][]{Meingast2019},
  the Pleiades \citep[cyan points; $(m - M)$ = 5.67, $A_V = 0.14$;][]{DR2HRD},
  and the subset of Pleiads with DR2 RVs (blue points).
    PARSEC isochrones with solar metallicity are overlaid in blue (130~Myr) and red (1~Gyr).
  The Pleiades sub-sample with RVs covers a nearly identical range in color and absolute magnitude as the stream's published membership. 
  This demonstrates that the stream's apparent turnoff color, which otherwise appears similar to NGC 6811's, 
  is biased by the lack of RV coverage in DR2 for warmer stars.
  The sole evolved member of the stream is 42~Cet, 
  which has an isochrone age of $\approx$1~Gyr; 
  this is at odds with our gyrochronology age 
  and suggests that it is an interloper.
  \textit{Right---}We identified 34 stars (red stars) that 
  are warmer than the \citet{Meingast2019} list (red open diamonds), and which closely track the Pleiades upper main sequence (blue points). Twenty two were found by pairing DR2 astrometry with literature RVs to determine their 3D kinematics, and the remaining twelve are co-moving neighbors in proper motion of \citet{Meingast2019} members. 
  Four of these candidates (red open stars) have peculiar abundances or are expected to have atypical photometry (open stars); 
  disregarding them, our high-mass candidates closely track the Pleiades upper main sequence.
      \label{f:cmd}}
\end{center}\end{figure*}

The absence of Psc--Eri members warmer than $\teff \approx$~7760~K on the main sequence would seem to favor an older age for the stream. However, as \citet{Meingast2019} pointed out, the stream lacks a clear main-sequence turnoff (MSTO). This is a problem: 
if the \stream\ is 1 Gyr old, there should be a well-defined MSTO (Figure~\ref{f:cmd} shows the case of NGC 6811). If the stream is young, the higher-mass stars should follow the Pleiades main sequence. Either way, these  stars should exist somewhere in the CMD, but they are either missing from the stream or missing from its membership catalog.

Identifying members of most star clusters is facilitated by their spatial overdensity and distance from Earth: proper motion are sufficient, and RVs are not strictly needed for finding candidate members. Identifying members of moving groups, stellar streams, and very nearby clusters (e.g., the Hyades) is more difficult because 3D kinematics are required. 
Accordingly, \citet{Meingast2019} used RVs to identify candidate \stream\ members. But the \textit{Gaia} Radial Velocity Spectrometer \citep{GaiaRVS, DR2RVS} provided measurements for stars with $3550 \lesssim \teff \lesssim 6900$~K  \citep{DR2RadialVelocities} in DR2. This data limitation means that the \citet{Meingast2019} criteria automatically precluded the identification of the MSTO for the \stream.

The left panel of Figure~\ref{f:cmd} plots the Pleiades membership \citep{DR2HRD}, and highlights those with DR2 RVs. The CMD for this Pleiades RV sample looks identical to the \citet{Meingast2019} membership for the \stream. We conclude that selecting members while requiring \textit{Gaia} RVs will exclude warmer members, if they exist (as well as the coolest, lowest-mass stars and hot white dwarfs).

\subsection{New, Massive Candidate Psc--Eri Members Support a Young Age} \label{s:hot}

If the \stream\ is the same age as the Pleiades, we should be able to identify hotter, more massive stars that are spatially and kinematically consistent with the \citet{Meingast2019} members. To this end, we queried DR2 for stars with $G < 10$ mag, $\gbr < 0.5$, and $M_G < 3$ mag, which returned 435,601 stars. We trimmed this to 6851 stars by selecting those consistent with the stream in R.A.~versus $\mu_\alpha \cos \delta$, R.A.~versus $\pi$, and decl. versus $\mu_\delta$ diagrams. Next, we searched SIMBAD \citep{simbad} for RV measurements for these stars. We found 2332 matches for which we could then calculate 3D galactic $UVW$ velocities.

Of these, 22 are within 5~\kms\ of the median value of the \citet{Meingast2019} members and within 20~pc of at least one member.\footnote{The median and maximum separation between nearest neighbors in the \citet{Meingast2019} membership list is 9 and 26~pc; the median and maximum velocity deviations from the stream's average $UVW$ velocity are 3 and 6~\kms.} While our velocity criterion is less restrictive than the 1.3~\kms\ velocity dispersion found by \citet{Meingast2019}, our larger threshold is justified by the fact that hotter, rapidly rotating stars will have less precise RVs than those for the FGK dwarfs reported in DR2.\footnote{Restricting the velocity criterion to $\leq$3~\kms\ reduces the 
candidate list from 22 to 11 stars.
Similarly, only 55\% of the \citet{Meingast2019} 
list has $\Delta v \leq$~3~\kms\ 
in Cartesian $UVW$ velocities.}

We also searched the 10~pc volume around every \citet{Meingast2019} member
for co-moving neighbors, according to the proper motion criterion 
$\Delta \mu < 2$~\mas, and found 377 co-moving candidates, including ten high-mass stars.
\citet{Oh2017} performed a similar exercise to identify co-moving pairs and larger groups 
using a more sophisticated 
algorithm applied to the \textit{Tycho--Gaia} Astrometric Solution \citep[TGAS;][]{TGAS} catalog, released with 
\textit{Gaia} DR1 \citep[see also][]{Andrews2017}. 
While the \stream\ was not identified as a co-moving system 
in their analysis,
they did identify seven high-mass stars as co-moving partners with 
members from \citet{Meingast2019}, adding two unique stars to our high-mass candidate list (12 co-moving neighbors in total).
Table~\ref{t:bcand} lists our 34 high-mass candidate members.

The right panel of Figure~\ref{f:cmd} shows the CMD for the Pleiades members together with the \citet{Meingast2019} \stream\ members and our high-mass candidate members.
The Pleiades has 239 members with DR2 RVs \citep{DR2HRD}, and \citet{Meingast2019} identified 256 members of the \stream. The sizes of these samples are approximately equal, so we expect that the \stream\ should have a similar number of higher-mass stars, and perhaps a similar population size and total mass.\footnote{For reference, the Pleiades has over 1000 known members: \citet{DR2HRD} identified 1332 members and  \citet{CG2018} identified 1061 members with \textit{Gaia} DR2 
\citep[see also][]{Sarro2014}. 
\citet{Adams2001} estimated a total mass of $\approx$800~\msun.} The Pleiades list has 43 more that are brighter and warmer than $\teff \approx 7000$~K RV cutoff, and we found 34 candidates in the stream.

Two of the five brightest candidates in the CMD 
are expected to have atypical photometry 
and should be excluded from an isochronal age analysis
\citep{Cummings2018young}: 
according to SIMBAD, 
$\lambda$~Tau is an Algol-type eclipsing binary and
omi~Aqr is a Be star. 
Focusing on the blue edge of the 
upper main sequence, 
the \stream\ is appears approximately 
coeval with the Pleiades.
The 80~Myr and 130~Myr PARSEC isochrones shown in 
in the right panel of Figure~\ref{f:cmd} 
do not diverge appreciably in color 
at the luminosities covered by the Pleiades and \stream\ samples. 
We postpone a precise isochronal analysis until we can 
validate the membership of our high-mass candidates with 
new RV measurements.

\subsection{How Did the Psc--Eri Stream Form?} \label{s:disp}
\citet{Meingast2019} estimated the \stream\ progenitor cluster mass to be $\approx$2000~\msun, noted that the Hyades initial mass has been estimated to be $\approx$1700~\msun, and concluded that since the Hyades still has a gravitationally bound core, the \stream, which has been dispersed, must be older. 

Indeed, if it were truly 1 Gyr old, the \stream\ would have had to be born as a dense cluster analogous to the Pleiades, Hyades, or NGC 6811, to survive for so long before disrupting. However, given that we now know that it is actually $\approx$120~Myr, this constraint on the stream's birth conditions is unnecessary. In their figure A.1 and table A.2, \citet{Meingast2019} identified four main clumps within the stream. These clumps are presently separated by $\approx$160~pc, and this clumpiness is similar to that seen in the much younger Tuc--Hor \citep{Kraus2014} or Sco--Cen associations \citep{Preibisch2008, Pecaut2016, WrightMama2018}, which are gravitationally unbound.

We suggest that the members of the \stream\ were not formed in a dense cluster but instead formed in a more decentralized fashion, similar to these OB associations.
If correct, this would resolve two challenges to our young age result: 
\begin{enumerate}
\item {\it Why does the stream not have a well defined core?} Our answer is that it never had one, but instead formed several smaller clumps.
\item {\it How could a 120-My-old cluster disperse its stars across 400~pc with such a low internal velocity dispersion?} The ends of stream had a head start, as they were born separated in space, and the members of each subgroup dispersed from there.
\end{enumerate}

According to the \citet{DR2HRD} membership lists, the Pleiades has 611 members within 5~pc of its center, and the Hyades has 195 members in the same size volume. In contrast, we suggest that the stream formed multiple approximately coeval clumps; therefore, each zero-age core density is much less than expected based on the present-day star count.

If we are correct, this would mean that the stream provides an environment to its stars that is distinct from that of the Pleiades, and which might be representative of a more common star formation channel in the Galaxy than 
dense cluster formation \citep[e.g.,][]{Clark2005}. That would make the \stream\ an excellent target for exoplanet searches, which have so far turned up nothing for the Pleiades \citep{Gaidos2017}.

\section{Conclusion} \label{s:end}

\citet{Meingast2019} discovered an exciting new stellar 
stream located relatively nearby ($d$ $\simeq$ 80-226 pc).
We were intrigued by its apparently old age ($\approx$1~Gyr), as this would make it a critical target for the calibration and validation of a variety of age-dating techniques, including stellar activity, rotation, lithium depletion, and other chemical clock techniques.

Using new time series photometry from \textit{TESS}, we measured \prot\ for 101 of the \stream's members. We found that the majority of these members actually overlap with the \prot\ distribution for the Pleaides, indicating that the \stream\ is only $\approx$120 Myr old.

By contrast to the CMD for the $\approx$1 Gyr old cluster NGC 6811, the \citet{Meingast2019} CMD for the \stream\ lacked a MSTO. We concluded that this is because the \stream\ is young, and that the more massive stars that would otherwise occupy the MSTO are warmer than the $\teff \lesssim 7000$~K cutoff for the \textit{Gaia} DR2 RV dataset; i.e., warmer stars could not be detected in DR2 as members by \citet{Meingast2019} because they lack 3D kinematics. We expanded the search for these missing members by pairing DR2 with RV measurements in the literature tabulated by SIMBAD, and also by searching for co-moving neighbors to the known members. We found 34 candidates that closely track the upper main sequence of the Pleiades, further strengthening our finding of a young age for the \stream. 

There is
one point on the \stream's CMD consistent with an old age: the evolved 42~Cet triple system. Given the indisputably young age for the \stream\ we found with gyrochronology, we suspect it is an interloper.

\citet{Meingast2019} estimated that the stream was formed with a total stellar mass similar to the Hyades. The Hyades has retained a dense cluster structure \citep[with tidal tails;][]{HyadesTail}, as has the Pleiades, while the stream is diffuse, with an elongated structure spanning 400~pc with four clumps. We argued that rather than being evidence for an older age, this structure indicates that the \stream's stars did not form in a dense cluster environment, but instead in the more decentralized fashion typical of OB associations.

If true, the \stream\ could become a valuable benchmark system for comparing environmental impact relative to the Pleiades, and for examining how photoevaporation sculpts planet sizes. To date, no planets have been found in the Pleiades \citep{Gaidos2017}.
The stream thus presents a new opportunity to search for Pleiades-aged planets.
Indeed, a Psc--Eri member has already been identified 
as a planet candidate host with \textit{TESS}.\footnote{First noted by Elisabeth Newton as a Psc--Eri member 
(priv.~comm.), TOI~451 
is a G dwarf with $\teff \approx 5530$~K (Gaia~DR2 4844691297067063424, CD$-$38~1467, TIC~257605131).
Our analysis of the \textit{TESS} 2~min light curves from Sectors 4 and 5 reveals two sets of transits, 
suggesting that TOI~451 hosts two planets with 
$P_{\rm orb, b} \approx  9.19$~d and
$P_{\rm orb, c} \approx 16.36$~d.
Follow-up efforts to rule out false positive scenarios and validate the planetary system are being coordinated by 
the \textit{TESS} Hunt for Young Moving group Exoplanets collaboration (THYME).} 

This is the first gyrochronology study using \textit{TESS} data, 
and it confirms that \textit{TESS} will be an exciting mission for stellar astrophysics. This is especially true given how \textit{TESS} records and releases FFI data. The existence of this stream was not known prior to the \textit{TESS} Cycle~1 call for proposals, and yet the FFI data were ready  for us to analyze immediately following the announcement of the stream's discovery by \citet{Meingast2019}. This is also the first time a stellar stream has been age-dated using gyrochronology, and our work demonstrates the potential for gyrochronology to serve as a powerful tool for Galactic archaeology.


\startlongtable
\begin{deluxetable*}{crcrcccrl}
\tablecaption{Rotation periods for \citet{Meingast2019} members of the \stream  \label{t:prot}}
\tablehead{\colhead{\#} & \colhead{\textit{Gaia} DR2 Source ID} & 
\colhead{R.A.} & \colhead{decl.} & 
\colhead{\gbr} & \colhead{$G$} & \colhead{$M_G$} & 
\colhead{\prot} & \colhead{Notes} \\
\colhead{} & \colhead{} & 
\colhead{[h:m:s]} & \colhead{[d:m:s]} & 
\colhead{[mag]} & \colhead{[mag]} & \colhead{[mag]} &
\colhead{[d]} & \colhead{}
}
\startdata
1 & 3198972700981234048 & 04:22:31.5 & $-$07:33:03.2 & 0.432 & 8.903 & 2.802 & 0.52 & Warm \\
2 & 5181474045115843072 & 03:10:47.3 & $-$06:34:29.8 & 0.446 & 8.562 & 2.954 & 0.87 & Warm \\
3 & 2516948215250061568 & 02:20:22.6 & +05:52:59.1 & 0.597 & 9.183 & 3.534 & 0.82 & Warm \\
4 & 3245408684793798528 & 04:02:15.4 & $-$05:53:48.2 & 0.604 & 9.425 & 3.513 & 0.56 & Conv. \\
5 & 6628071944405827712 & 22:36:31.1 & $-$21:35:06.0 & 0.647 & 8.967 & 3.835 & 0.94 & Conv. \\
6 & 2988966044497883392 & 05:22:51.9 & $-$11:47:47.8 & 0.648 & 10.345 & 3.688 & 0.79 & Conv. \\
7 & 2456987757379368064 & 01:32:34.4 & $-$12:51:09.7 & 0.654 & 9.043 & 3.865 & 1.24 & Conv. \\
8 & 3186195241994234880 & 04:43:02.6 & $-$07:53:54.6 & 0.655 & 9.999 & 3.766 & 1.71 & Conv. \\
9 & 2988096919213031808 & 05:02:35.2 & $-$12:31:20.4 & 0.661 & 10.280 & 3.869 & 0.91 & Conv. \\
10 & 2987729922847457280 & 05:07:09.2 & $-$13:34:07.7 & 0.668 & 10.156 & 3.896 & 0.97 & Conv. \\
11 & 3204844780267292288 & 04:29:21.6 & $-$02:49:47.1 & 0.673 & 9.760 & 4.020 & 1.98 & Conv. \\
12 & 2405544971274027904 & 23:21:22.3 & $-$17:30:58.5 & 0.680 & 9.222 & 4.024 & 1.44 & Conv. \\
13 & 3190206672727634816 & 04:01:28.8 & $-$11:19:25.7 & 0.682 & 9.920 & 4.059 & 2.26 & Conv. \\
14 & 5182223980765557248 & 03:18:22.8 & $-$04:29:29.0 & 0.687 & 9.788 & 4.054 & 1.29 & Conv. \\
15 & 2982998926174605824 & 05:10:30.1 & $-$16:08:04.1 & 0.691 & 10.646 & 3.970 & 1.70 & Conv. \\
16 & 2492898356897645184 & 02:18:04.2 & $-$03:50:14.4 & 0.700 & 9.453 & 4.126 & 2.67 & Conv. \\
17 & 3256702490277205376 & 04:03:24.9 & $-$00:46:45.2 & 0.701 & 9.928 & 4.180 & 3.08 & Conv. \\
18 & 5104477754084350464 & 03:15:18.8 & $-$17:56:36.4 & 0.731 & 9.606 & 4.279 & 2.55 & Conv. \\
19 & 5147686052794315904 & 02:02:10.9 & $-$16:34:03.4 & 0.746 & 9.643 & 4.410 & 2.53 & Conv. \\
20 & 3197608241410937216 & 04:32:01.8 & $-$08:53:13.7 & 0.758 & 10.494 & 4.476 & 2.85 & Conv. \\
21 & 2489889607752127360 & 02:18:43.9 & $-$04:00:56.0 & 0.759 & 10.019 & 4.390 & 2.39 & Conv. \\
22 & 2346216668164370432 & 00:54:13.5 & $-$22:53:07.8 & 0.764 & 9.441 & 4.452 & 3.10 & Conv. \\
23 & 5129126330877050240 & 02:46:34.6 & $-$18:54:17.5 & 0.766 & 9.784 & 4.520 & 3.00 & Conv. \\
24 & 5070969209513725568 & 02:38:36.5 & $-$25:15:07.6 & 0.775 & 9.435 & 4.462 & 5.69 & Slow \\
25 & 2493286445846897664 & 02:15:46.4 & $-$02:36:32.5 & 0.821 & 9.916 & 4.813 & 3.84 & Conv. \\
26 & 3197753548744455168 & 04:33:55.4 & $-$08:19:27.9 & 0.827 & 10.892 & 4.799 & 3.96 & Conv. \\
27 & 2513568007268649728 & 02:14:47.2 & +02:14:20.4 & 0.865 & 10.473 & 5.025 & 4.30 & Conv. \\
28 & 5179904664065847040 & 03:09:03.7 & $-$07:03:55.8 & 0.873 & 10.382 & 4.979 & 4.78 & Conv. \\
29 & 5168681021169216896 & 03:29:30.3 & $-$07:10:13.8 & 0.895 & 10.827 & 5.204 & 5.32 & Conv. \\
30 & 3253302456727341696 & 04:07:34.7 & $-$02:04:33.2 & 0.901 & 10.982 & 5.112 & 4.83 & Conv. \\
31 & 3176016268285396864 & 04:21:35.2 & $-$14:01:29.9 & 0.902 & 11.335 & 5.263 & 5.11 & Conv. \\
32 & 2531732317316926336 & 01:15:31.7 & $-$02:50:46.4 & 0.903 & 10.841 & 5.210 & 4.87 & Conv. \\
33 & 2495781619982992640 & 02:45:01.2 & $-$02:25:46.3 & 0.921 & 10.780 & 5.304 & 5.00 & Conv. \\
34 & 2968825259219765120 & 05:29:28.5 & $-$19:17:58.8 & 0.924 & 12.032 & 5.264 & 5.22 & Conv. \\
35\tablenotemark{a} & 4844691297067063424 & 04:11:51.9 & $-$37:56:23.03 & 0.927 & 10.750 & 5.280 & 5.02 & Conv. \\
36 & 2496200774431287424 & 02:30:58.8 & $-$03:03:04.9 & 0.928 & 10.415 & 5.328 & 5.45 & Conv. \\
37 & 4980826504625538048 & 00:38:12.2 & $-$43:00:24.8 & 0.935 & 10.317 & 5.401 & 5.24 & Conv. \\
38 & 4842810376267950464 & 03:47:56.3 & $-$41:56:24.9 & 0.936 & 10.762 & 5.383 & 5.70 & Conv. \\
39 & 3198734278756825856 & 04:26:27.1 & $-$07:39:39.7 & 0.966 & 11.469 & 5.445 & 8.35 & Slow \\
40 & 5045955865443216640 & 03:00:46.9 & $-$37:08:01.5 & 0.976 & 10.323 & 5.363 & 3.90 & Rapid \\
41 & 3193528950192619648 & 03:57:04.0 & $-$10:14:00.9 & 0.994 & 11.297 & 5.534 & 5.54 & Conv. \\
42 & 3187547465200970368 & 04:46:12.3 & $-$07:32:24.4 & 0.997 & 11.733 & 5.519 & 5.43 & Conv. \\
43 & 3245140743257978496 & 03:54:01.0 & $-$06:14:14.6 & 1.017 & 11.146 & 5.434 & 5.66 & Conv. \\
44 & 3205573756476323328 & 04:23:54.6 & $-$02:33:43.4 & 1.029 & 11.604 & 5.634 & 6.05 & Conv. \\
45 & 3185678437170300800 & 04:34:42.8 & $-$08:57:18.5 & 1.052 & 11.899 & 5.796 & 0.55 & Rapid \\
46 & 5103353606523787008 & 03:18:03.8 & $-$19:44:14.2 & 1.058 & 10.473 & 5.227 & 1.26 & Rapid \\
47 & 3187477818011309568 & 04:47:58.2 & $-$07:49:25.2 & 1.059 & 11.994 & 5.818 & 7.02 & Conv. \\
48 & 3009905594911137664 & 05:26:30.0 & $-$12:01:21.1 & 1.073 & 12.576 & 5.905 & 6.50 & Conv. \\
49 & 5083255496041631616 & 03:57:35.1 & $-$24:28:42.2 & 1.077 & 11.131 & 5.885 & 12.22 & Slow \\
50 & 3243665031151732864 & 03:48:38.3 & $-$06:41:52.6 & 1.082 & 11.460 & 5.953 & 6.84 & Conv. \\
51 & 3192643431015406464 & 04:21:53.2 & $-$08:43:16.1 & 1.088 & 11.752 & 5.873 & 6.84 & Conv. \\
52 & 5029398079322118912 & 01:13:42.4 & $-$31:11:39.6 & 1.088 & 10.804 & 5.959 & 6.64 & Conv. \\
53 & 2596395760081700608 & 22:39:53.5 & $-$16:36:23.3 & 1.097 & 11.508 & 5.994 & 6.97 & Conv. \\
54 & 2979827384884386176 & 04:58:02.5 & $-$17:10:27.7 & 1.113 & 12.317 & 6.033 & 6.34 & Conv. \\
55 & 3196687812738993152 & 04:06:00.2 & $-$06:53:50.0 & 1.128 & 11.842 & 6.073 & 7.02 & Conv. \\
56 & 2491594263092190464 & 02:10:22.3 & $-$03:50:56.7 & 1.136 & 11.533 & 6.153 & 2.26 & Rapid \\
57 & 2402197409339616768 & 22:39:01.4 & $-$18:52:55.7 & 1.142 & 11.219 & 6.155 & 7.80 & Conv. \\
58 & 3013355999838366336 & 05:25:14.6 & $-$10:25:49.4 & 1.159 & 12.806 & 6.185 & 7.11 & Conv. \\
59 & 5096891158212909312 & 04:12:46.0 & $-$16:19:29.1 & 1.181 & 11.994 & 6.187 & 3.68 & Rapid \\
60 & 4871041608622321664 & 04:28:28.9 & $-$33:53:45.1 & 1.193 & 11.630 & 6.037 & 6.50 & Rapid \\
61 & 7324465427953664 & 03:05:14.1 & +06:08:53.5 & 1.197 & 12.043 & 6.247 & 4.40 & Rapid \\
62 & 5097262136011410944 & 03:58:54.7 & $-$17:05:53.2 & 1.199 & 11.649 & 6.311 & 7.58 & Conv. \\
63 & 2418664520110763520 & 23:49:55.1 & $-$15:43:42.0 & 1.213 & 11.607 & 6.396 & 2.94 & Rapid \\
64 & 5179037454333642240 & 02:39:10.9 & $-$05:32:22.5 & 1.215 & 11.765 & 6.373 & 6.42 & Rapid \\
65 & 2484875735945832704 & 01:24:24.7 & $-$03:16:39.0 & 1.222 & 11.791 & 6.398 & 8.40 & Conv. \\
66 & 2393862836322877952 & 23:40:37.5 & $-$18:11:37.9 & 1.239 & 11.485 & 6.472 & 7.70 & Conv. \\
67 & 2594993646533642496 & 22:31:13.9 & $-$17:04:52.4 & 1.242 & 11.934 & 6.451 & 6.10 & Rapid \\
68 & 3199896668704440064 & 04:38:55.5 & $-$06:40:25.0 & 1.242 & 12.458 & 6.305 & 4.28 & Rapid \\
69 & 5114686272872474880 & 03:47:25.8 & $-$12:32:30.9 & 1.247 & 12.634 & 6.550 & 9.00 & Conv. \\
70 & 2433715455609798784 & 23:36:52.1 & $-$11:25:01.7 & 1.249 & 11.737 & 6.442 & 6.30 & Rapid \\
71 & 5106733402188456320 & 03:24:25.2 & $-$15:50:05.4 & 1.278 & 11.517 & 6.197 & 0.62 & Rapid \\
72 & 2390974419276875776 & 23:48:32.4 & $-$18:32:57.4 & 1.283 & 11.583 & 6.618 & 6.50 & Rapid \\
73 & 3172630287868034944 & 04:26:48.2 & $-$15:25:47.4 & 1.307 & 12.334 & 6.599 & 5.85 & LM \\
74 & 5161117923061794688 & 02:59:52.0 & $-$09:47:35.8 & 1.308 & 12.063 & 6.639 & 5.45 & LM \\
75 & 5155187986271622912 & 03:20:33.3 & $-$14:16:58.4 & 1.320 & 12.307 & 6.687 & 1.93 & LM \\
76 & 5129876953722430208 & 02:29:28.5 & $-$20:12:16.8 & 1.334 & 11.695 & 6.739 & 8.00 & LM \\
77 & 2339984636258635136 & 23:56:53.7 & $-$23:17:24.6 & 1.349 & 11.475 & 6.765 & 8.35 & LM \\
78 & 3195826963854173056 & 04:06:29.4 & $-$07:35:32.2 & 1.366 & 12.987 & 6.797 & 10.73 & Slow \\
79 & 2349094158814399104 & 00:47:18.0 & $-$22:45:08.1 & 1.366 & 10.929 & 6.051 & 1.30 & LM \\
80 & 3247412647814482816 & 03:32:30.9 & $-$06:13:09.1 & 1.382 & 12.327 & 6.898 & 6.66 & LM \\
81 & 4975223840046231424 & 00:47:38.5 & $-$47:41:45.8 & 1.420 & 11.514 & 6.985 & 5.80 & LM \\
82 & 3191365111308746880 & 04:24:30.4 & $-$10:41:02.4 & 1.423 & 12.901 & 6.944 & 7.02 & LM \\
83 & 3197607794734344320 & 04:31:51.6 & $-$08:54:03.5 & 1.439 & 13.133 & 7.059 & 8.02 & LM \\
84 & 3171136944919260928 & 04:36:36.1 & $-$17:47:23.7 & 1.439 & 12.879 & 6.940 & 7.85 & LM \\
85 & 3206907086126334464 & 05:13:25.5 & $-$08:19:52.2 & 1.465 & 13.320 & 7.086 & 6.66 & LM \\
86 & 3177883999240571904 & 04:35:35.3 & $-$12:47:47.6 & 1.480 & 12.828 & 6.356 & 4.00 & LM \\
87 & 2488721720245150336 & 02:26:53.3 & $-$05:17:45.2 & 1.498 & 12.485 & 7.198 & 5.20 & LM \\
88 & 5159567164990031360 & 03:04:46.0 & $-$12:16:57.9 & 1.560 & 12.613 & 7.098 & 0.45 & LM \\
89 & 5081912751826042624 & 03:47:01.6 & $-$26:16:11.2 & 1.577 & 12.968 & 7.389 & 4.50 & LM \\
90 & 5118895478259982336 & 02:26:07.0 & $-$24:54:49.0 & 1.579 & 11.728 & 6.658 & 5.61 & LM \\
91 & 5117016378528360448 & 02:17:14.6 & $-$27:16:41.9 & 1.591 & 12.551 & 7.451 & 5.92 & LM \\
92 & 5149427640557882368 & 02:02:58.3 & $-$13:37:46.8 & 1.594 & 11.982 & 7.442 & 11.96 & Slow \\
93 & 4832163770817481856 & 03:58:17.4 & $-$46:34:13.0 & 1.630 & 12.992 & 7.559 & 7.80 & LM \\
94 & 2531488844210764544 & 01:08:57.0 & $-$03:01:32.0 & 1.661 & 13.074 & 7.612 & 2.26 & LM \\
95 & 2480756793589426944 & 01:33:49.3 & $-$04:28:41.7 & 1.670 & 13.216 & 7.682 & 6.50 & LM \\
96 & 2355466790769878400 & 00:55:21.5 & $-$21:24:03.7 & 1.689 & 12.341 & 7.578 & 6.84 & LM \\
97 & 5114516020369038848 & 03:51:15.9 & $-$12:23:46.4 & 1.732 & 13.375 & 7.484 & 0.68 & LM \\
98 & 5068272932125221504 & 02:26:04.6 & $-$29:23:48.9 & 1.768 & 12.815 & 8.000 & 7.80 & LM \\
99 & 5094664333632217088 & 04:02:18.0 & $-$18:42:45.4 & 1.771 & 12.631 & 7.134 & 2.62 & LM \\
100 & 5121805541941481472 & 01:57:17.2 & $-$25:13:49.6 & 1.784 & 12.793 & 7.931 & 5.45 & LM \\
101 & 4984094970441940864 & 01:21:49.7 & $-$42:01:22.3 & 1.852 & 12.731 & 8.099 & 5.45 & LM \\
\enddata
\tablecomments{Columns: \# is the row number sorted by \gbr;
R.A., decl., \gbr\, $G$, and $M_G = G - 5\,\log_{10} (100/\pi)$ are from \textit{Gaia} DR2;
\prot\ is measured from \textit{TESS} FFI data (days). 
The notes indicate if a star is 
converged on the slow sequence (``Conv.''), slower than the converged sequence (``Slow''), 
more rapid than the converged sequence (``Rapid''), 
has a lower mass (``LM'') than the converged sequence limit, 
or is too warm to efficiently spin down (``Warm'').
notes on particular stars.
Six stars appear to rotate more slowly than the bulk of the sample.
Blending is not a concern for these stars 
(i.e., none have bright neighbors in DR2 within 1\farcm5), 
their spot-modulated light curves show unambiguous periodicity, 
and they do not appear to be binaries according to their
photometry, RV errors ($\sigma < 2$~\kms), and kinematics.
It is unclear to us why they are outliers.
}
\tablenotetext{a}{This star has been identified as a planet candidate host by \textit{TESS} (TOI~451, TIC~257605131), 
and appears to show two sets of transits with periods of 9.19~d and 16.36~d, 
which await validation.}
\end{deluxetable*}


\begin{deluxetable*}{rccrDDDDccDl}
\tablecaption{Candidate massive members of the \stream} \label{t:bcand}
\tablehead{\colhead{\#} & \colhead{\textit{Gaia} DR2 ID} & 
\colhead{$\alpha$(ICRS)} & \colhead{$\delta$(ICRS)} & 
\multicolumn2c{$G_{\rm BP} - G_{\rm RP}$} 
& \multicolumn2c{$G$} & \multicolumn2c{$M_G$} & 
\multicolumn2c{RV}  & \colhead{$\Delta v$} & \colhead{$\Delta \mu$} & \multicolumn2c{$\Delta r$} & \colhead{Name} \\
\colhead{} & \colhead{} & 
\colhead{[h:m:s]} & \colhead{[d:m:s]} & 
\multicolumn2c{[mag]} 
& \multicolumn2c{[mag]} & \multicolumn2c{[mag]} & 
\multicolumn2c{[\kms]}  & \colhead{[\kms]} & \colhead{[\mas]} & \multicolumn2c{[pc]} & \colhead{}
}
\decimals
\startdata
1 & 3305012316783145728 & 04:00:40.81 &   +12:29:25.0  & $-$0.008 & 3.387 & $-$2.039 & 17.8 & 1.3 & 2.3 & 18.7 & $\lambda$\,Tau\tablenotemark{a} \\
2 & 2676509823708845056 & 22:03:18.87 & $-$02:09:19.5  & $-$0.083 & 4.622 & $-$1.178 & 11.0 & 3.4 & 1.1 & 17.7 & o\,Aqr\tablenotemark{b} \\ 
3 & 5021010046848175616 & 02:04:29.45 & $-$29:17:48.3  & $-$0.193 & 4.638 & $-$0.447 & 18.5 & 4.1 & 5.8 & 6.3 & $\nu$\,For\tablenotemark{c} \\
4 & 2391220091406075648 & 23:44:12.11 & $-$18:16:37.1  & $-$0.080 & 5.185 & $-$0.102 & 16.0 & 1.8 & 3.0 & 6.9 &  106\,Aqr \\
5 & 2390144081839340288 & 23:51:21.37 & $-$18:54:33.1  & $-$0.153 & 5.125 &    0.096 & 12.7 & 2.9 & 0.3 & 0.9 & 108\,Aqr\tablenotemark{d} \\ 
6 & 2597327566122330880 & 22:47:42.80 & $-$14:03:23.3  & $-$0.037 & 5.666 & 0.481 & 15.0 & 3.9 & 1.19 & 11.4 & $\tau^1$\,Aqr \\
7 & 2734781844037454592 & 22:24:00.51 &   +15:16:53.1  & $-$0.085 & 6.764 & 0.503 & 5.0 & 2.8 & 0.9 & 14.6 & HD\,212442 \\
8 & 2428341184508675456 & 00:14:54.54 & $-$09:34:10.6  & $-$0.103 & 5.747 & 0.531 & 19.9 & 3.9 & 0.7 & 11.8 & HR\,51 \\ 
9 & 4878579825983429248 & 04:36:50.91 & $-$30:43:00.3  & $-$0.100 & 6.252 & 0.569 & 14.5 & 4.1 & 2.3 & 10.2 & HR\,1476 \\ 
10 & 2746298781663140352 & 23:55:07.82 &  +07:04:15.2  & $-$0.074 & 6.196 & 0.762 & 16.8 & 4.1 & 2.4 & 4.5 & 26\,Psc \\ 
11 & 6549670305714644608 & 23:06:53.67 & $-$38:53:32.2 &    0.054 & 5.631 & 0.921 & 11.9 & 3.1 & 2.0 & 17.5 & $\upsilon$\,Gru\tablenotemark{e} \\ 
12 & 5045432364765457792 & 02:57:32.63 & $-$38:11:27.2 & $-$0.022 & 6.395 & 1.062 & 19.6 & 2.0 & 1.7 & 18.6 &  HR\,893 \\ 
13 & 2410222091875449216 & 23:09:49.58 & $-$14:30:38.1 & 0.033 & 6.411 & 1.276    & 16.6 & 3.3 & 0.9 & 7.5 & HR\,8816  \\ 
14 & 3252923090855768064 & 04:04:53.38 & $-$02:25:37.8 & $-$0.007 & 7.056 & 1.353 & $\cdots$ & $\cdots$ & 1.4 & 5.1 & HD\,25752 \\ 
15 & 5175455696422545664 & 02:35:24.47 & $-$09:21:02.8 & 0.016 & 7.097 & 1.445  & 19.1 & 1.6 & 3.4 & 8.8 & HD\,16152 \\  
16 & 3192744139408470272 & 04:21:35.19 & $-$08:06:31.1 & 0.022 & 7.483 & 1.537 & $\cdots$ & $\cdots$ & 1.5 & 5.0 & HD\,27665 \\ 
17 & 2982108287397220992 & 05:23:07.86 & $-$17:13:26.1 & 0.064 & 8.250 & 1.567 & $\cdots$ & $\cdots$ & 0.7 & 6.2 & HD\,35308  \\ 
18 & 2741090498161113344 & 00:15:57.32 &   +04:15:03.8 & 0.052 & 7.107 & 1.607 & 12.6 & 2.1 & 1.3 & 6.2 & HD\,1160\,A\tablenotemark{f} \\ 
19 & 2697317256631380736 & 21:58:36.60 &   +06:00:49.8 & 0.083 & 7.964 & 1.679 & $\cdots$ & $\cdots$ & 1.5 & 4.6 & HD\,208800 \\ 
20 & 2721809496615333248 & 22:00:50.95 &   +07:51:08.5 & 0.070 & 7.989 & 1.713 & $\cdots$ & $\cdots$ & 1.2 & 3.8 & HD\,209105 \\ 
21 & 2986248601510045184 & 05:00:01.23 & $-$15:47:55.2 & 0.060 & 8.372 & 1.732 & 14.9 & 4.7 & 1.1 & 6.6 & HD\,32077  \\ 
22 & 2542373597009506944 & 00:31:40.77 & $-$01:47:37.4 & 0.094 & 7.041 & 1.776 & $\cdots$ & $\cdots$ & 1.0 & 5.2 & HD\,2830  \\ 
23 & 2971453886581625600 & 05:36:04.99 & $-$16:51:45.1 & 0.105 & 8.561 & 1.783 & $\cdots$ & $\cdots$ & 0.8 & 11.5 & HD\,37190 \\ 
24 & 2391000395239148672 & 23:52:39.91 & $-$18:33:42.8 & 0.122 & 6.804 & 1.806 & 13.0 & 2.9 & 0.9 & 0.1 & HD\,223785 \\ 
25 & 5127759431765387392 & 02:45:18.39 & $-$20:24:05.9 & 0.058 & 7.106 & 1.827 & $\cdots$ & $\cdots$ & 0.3 & 3.1 & HD\,17224  \\ 
26 & 2982652206352410624 & 05:12:33.00 & $-$17:27:16.5 & 0.117 & 8.514 & 1.899 & $\cdots$ & $\cdots$ & 1.8 & 6.2 & HIP\,2427 \\ 
27 & 2982652275071888128 & 05:12:29.72 & $-$17:27:08.9 & 0.119 & 8.564 & 1.961 & $\cdots$ & $\cdots$ & 0.7 & 6.7 & HD\,33857 \\ 
28 & 3182650382147268992 & 05:07:44.10 & $-$09:51:53.5 & 0.212 & 8.717 & 2.137 & 13.9 & 3.5 & 0.5 & 13.3 & HD\,33126\tablenotemark{e}  \\ 
29 & 5145324782854148096 & 02:13:19.11 & $-$14:54:27.3 & 0.495 & 7.556 & 2.226 & 19.0 & 3.1 & 9.3 & 8.1 & HD\,13722 \\ 
30 & 3185719600136840064 & 04:35:06.82 & $-$08:41:36.6 & 0.273 & 8.538 & 2.327 & $\cdots$ & $\cdots$ & 1.3 & 6.7 & HD\,29152 \\ 
31 & 3300937801567693824 & 04:15:00.92 &   +10:44:53.1 & 0.283 & 7.652 & 2.385 & 16.0 & 2.1 & 3.1 & 18.6 & HD\,26843 \\ 
32 & 2736194815262723712 & 22:34:06.27 &   +16:01:27.2 & 0.318 & 8.886 & 2.549 & 8.2 & 2.5 & 0.4 & 14.7 & HD\,213838 \\ 
33 & 5155416822128110208 & 03:18:13.96 & $-$13:49:45.4 & 0.303 & 8.226 & 2.625 & $\cdots$ & $\cdots$ & 1.7 & 2.0 & HD\,20573 \\ 
34 & 3175513589608066048 & 04:19:34.07 & $-$15:10:11.8 & 0.415 & 8.999 & 2.927 & 19.5 & 2.6 & 2.8 & 3.6 & HD\,27467\tablenotemark{e} \\ 
\enddata
\tablecomments{Columns---\# is the row number, sorted by $M_G$;
$\alpha$(ICRS, epoch 2015.5), $\delta$(ICRS, epoch 2015.5), 
\gbr, $G$, and $M_G = G - 5\,\log_{10} (100/\varpi)$ from \textit{Gaia} DR2;
radial velocity obtained from SIMBAD;
$\Delta v$ is the absolute deviation of $UVW$ velocities from the stream's median value (\kms);
$\Delta \mu$ is minimum difference in proper motion relative to the nearest neighbor in the \citet{Meingast2019} list (\mas); 
$\Delta XYZ$: the physical distance (pc) to the nearest \citet{Meingast2019} member; 
common aliases.
Notes on particular stars from SIMBAD are provided 
below.
}
\tablenotetext{a}{Algol-type EB}
\tablenotetext{b}{Be star}
\tablenotetext{c}{$\alpha^2$ CVn variable}
\tablenotetext{d}{Peculiar composition}
\tablenotetext{e}{Binary or multiple star}
\tablenotetext{f}{HD~1160 has two low-mass companions \citep{Nielsen2012}---HD~1160\,C is an M3.5 dwarf (Gaia DR2 2741090498159705216), 
and HD~1160\,B is a brown dwarf candidate with an estimated mass of 
39-166~\mjup\ \citep{Maire2016}, 35-90~\mjup, and 70-90~\mjup\ \citep{Garcia2017}, 
depending on the age of the host star.
Interpolating the $125 \pm 15$~Myr evolutionary models from \citet{Baraffe2015} at the 
\citet{Garcia2017} temperature ($\teff = 3050 \pm 50$~K) and 
luminosity, corrected with the \textit{Gaia} DR2 parallax ($\log L/L_\odot = -2.59 \pm 0.05$~dex,
we infer a mass $M_B =  0.12 \pm 0.01$~\msun\ ($\approx$123~\mjup).
This is greater than the hydrogen-burning limit 
and indicates that HD~1160\,B is probably a very-low-mass star and not a brown dwarf.}
\end{deluxetable*}

\acknowledgments

We thank Stefan~Meingast for kindly providing us with 
the \citet{Meingast2019} membership list prior to 
its posting to the CDS, and Tim White for helpful discussions about 42~Ceti.
The association of TOI~451 with Psc--Eri was first noted 
by Elisabeth Newton; we thank her and the THYME collaboration, 
including Aaron Rizzuto, 
Andrew Vanderburg, Andrew Mann, and Benjamin Tofflemire, 
and Adam Kraus for discussing this exciting planet candidate with us.
    
J.L.C. is supported by the National Science Foundation 
Astronomy and Astrophysics Postdoctoral Fellowship under award AST-1602662.

Part of this research was carried out at the Jet Propulsion Laboratory, 
California Institute of Technology, under a contract with NASA.

The Center for Exoplanets and Habitable Worlds is supported by the
Pennsylvania State University, the Eberly College of Science, and the
Pennsylvania Space Grant Consortium.

This work has made use of data from the European Space Agency (ESA)
mission {\it Gaia},\footnote{\url{https://www.cosmos.esa.int/gaia}} 
processed by the {\it Gaia} Data Processing and Analysis Consortium 
(DPAC).\footnote{\url{https://www.cosmos.esa.int/web/gaia/dpac/consortium}} 
Funding for the DPAC has been provided by national institutions, in particular
the institutions participating in the {\it Gaia} Multilateral Agreement.

This research made use of NASA's Astrophysics Data System, 
and the VizieR and SIMBAD \citep{simbad} databases, operated at CDS, Strasbourg, France.

\vspace{5mm}
\facilities{\textit{TESS}, \textit{Gaia}}


\software{The IDL Astronomy User's Library \citep{IDLastro}\footnote{\url{https://github.com/wlandsman/IDLAstro}}, TESScut \citep[i.e., Astrocut;][]{Astrocut}\footnote{\url{https://mast.stsci.edu/tesscut}}
}

\bibliographystyle{aasjournal}

\end{document}